\documentclass[%
 aip,
 amsmath,amssymb,
 reprint,%
]{revtex4-1}

\usepackage{graphicx}
\usepackage{subfigure}
\usepackage{dcolumn}
\usepackage{bm}

\usepackage[utf8]{inputenc}
\usepackage[T1]{fontenc}
\usepackage{mathptmx}
\usepackage{color}

\usepackage{fancyhdr}
\begin{document}

\preprint{AP/123-QED}

\title{Delay Induced Swarm Pattern Bifurcations in Mixed Reality Experiments}

\author{Victoria Edwards}
\email{victoria.edwards@nrl.navy.mil}
\affiliation{Navy Center for Applied Research in Artificial Intelligence, Naval Research Laboratory, Washington, DC, 20375, USA}

\author{Philip deZonia}
\affiliation{Mechanical Engineering and Applied Mechanics, University of Pennsylvania, Philadelphia, Pennsylvania 19104, USA}

\author{M. Ani Hsieh}%
 \affiliation{Mechanical Engineering and Applied Mechanics, University of Pennsylvania, Philadelphia, Pennsylvania 19104, USA}%

 \author{Jason Hindes}
 \affiliation{Nonlinear Dynamical Systems, Naval Research Laboratory, Washington, DC, 20375, USA}

 \author{Ioana Triandaf}
 \affiliation{Nonlinear Dynamical Systems, Naval Research Laboratory, Washington, DC, 20375, USA}

\author{Ira B Schwartz}
\affiliation{Nonlinear Dynamical Systems, Naval Research Laboratory, Washington, DC, 20375, USA}%

\date{\today}

\begin{abstract}
Swarms of coupled mobile agents subject to inter-agent wireless communication delays are known to exhibit multiple dynamic patterns in space that depend on the strength of the interactions and the magnitude of the communication delays. 
We experimentally demonstrate communication delay-induced bifurcations in the spatio-temporal patterns of robot swarms using two distinct hardware platforms in a mixed reality framework.  
Additionally, we make steps toward experimentally validating theoretically predicted parameter regions where transitions between swarm patterns occur. 
We show that multiple rotation patterns persist even when collision-avoidance strategies are incorporated, and we show the existence of multi-stable, co-existing rotational patterns not predicted by usual mean field dynamics.
Our experiments are the first significant steps towards validating existing theory and the existence and robustness of the delay-induced patterns in real robotic swarms.
\end{abstract}

\maketitle
\thispagestyle{fancy}   
\begin{quotation}
There is interest in deploying swarms of autonomously interacting robots for cooperative tasks such as search and rescue, exploration and mapping, distributed sensing and estimation, and more.
However, accounting for the uncertainties of the physical world is challenging when validating with simulation alone, and experimental validation using large numbers of real robots requires overcoming significant logistical and resource constraints; e.g., space, cost, manpower. 
A useful intermediate approach for systematic experimental validation is to couple simulation with real robots in a mixed reality framework.
Mixed reality retains the key features of physical experiments that are hard to capture through simulation alone.
In particular, realistic inter-agent wireless communication is extremely difficult to capture in simulation given the computational complexities of faithfully modeling the physics of RF propagation.
The mixed reality framework enables the study of the impact of wireless communication effects, such as delays in communication, on swarms with physical robots.
Existing theory has shown that when inter-agent communication delays are introduced into an ideal swarm with simple interaction rules, collective rotation patterns can emerge.
In this work, we take steps toward testing such patterns with swarms of real robots, and characterize when transitions occur between patterns in a mixed reality setting.
\end{quotation}

\section{Introduction}
Natural swarms have inspired researchers to understand how simple organisms produce complex emergent patterns and behaviors.
Swarms in nature are composed of individual agents with relatively simple behaviors that interact locally to synergistically yield complex collective behavior. 
Examples of such swarms in nature include: schools of fish \cite{Couzin2013, Calovi2014}, flocks of starlings \cite{Leonard2013, Cavagna2015} and jackdaws \cite{Ouellette19}, colonies of bees \cite{Li_Sayed_2012} and ants \cite{Theraulaz2002}, aggregation of locust \cite{Topaz12}, and crowds of people \cite{Rio_Warren_2014}.

When examining biological swarms, recent analysis has shown that there exists a delay in reaction time between agents.  In other words, as agents move, they react in response to the past positions of their neighbors rather than their instantaneously detected positions. 
For example, delays have been measured in schooling fish~\cite{JL_fish}, bats~\cite{Luca_bats}, birds~\cite{Nagy_pigeons} and crowds of people~\cite{JF_people}.   
Since natural agents move in an almost continuous manner, it is natural to model swarms based on biological ideas as continuous systems with communication delays. 

Swarms can be modeled by differential delay equations where the delay is included in the communication network between agents.
Such delays can act as a destabilizing parameter, which means for a swarm with communication delay different spatio-temporal patterns
will emerge.
Pattern emergence was shown by analyzing the mean field of a globally coupled swarm in the presence of communication delays where a Hopf bifurcation can be observed \cite{Romero2012}.
Recent work has generalized this result with an exact stability analysis and the inclusion of range dependent delay \cite{Hindes2020, Schwartz2020}.
In Szwaykowska et al. \cite{Szwaykowska2016}, theoretical analysis showed that the delay-induced bifurcation picture is robust to random link removal in a swarm's communication network and to agent heterogeneity.
Preliminary mixed reality experiments tested one of the theoretically predicted swarming behaviors for a restricted parameter set, but did not show transitions between swarm behaviors based on parameter changes.

In general, existing works in the design and control of artificial swarms have focused on the synthesis of local interaction rules that give rise to global swarm behavior.
These works focus on bottom-up strategies where the objective is to develop provably correct single robot strategies that yield some desired swarm behavior \cite{Tanner03a, Tanner03b, Tanner07, Jadbabaie03, Viraghn14, Gazi05, Desai01}. 
Unsurprisingly, these works are based on a strict set of assumptions, which are necessary to ensure the desired emergent behavior. 
Nevertheless, while the strategies have been tested in simulation, validation on physical systems is often problematic since the necessary assumptions do not always hold in the real-world.  
Thus, the extensive body of work has provided a strong theoretical foundation for the design of swarming strategies but few have been experimentally validated on actual systems.

Experimental validation with physical robots invariably introduces uncertainty in the form of actuation, sensing, and robot-robot and robot-environment interaction noise.
Since swarms are complex nonlinear dynamical systems, they can typically exhibit multiple steady-state patterns \cite{Romero2014}.
In the presence of noise it is possible to transition from one steady-state to another \cite{Freidlin1998, Szwaykowska2018}.
Additionally, changes in system parameters can result in changes to the stability nature of steady-states \cite{Wells15, Hartle17, Ansmann16}.
In this work we conduct experiments with robots, which requires dealing with uncertainties inherent in physical experiments as well as changes in system parameters. 

 The dramatic reduction in the price to performance ratio of embedded processors, sensors, and computers and the ubiquity of wireless communications technologies have made experiments with ever larger number of robots more viable.  
Examples of these experiments include the Kilobots \cite{Rubenstein12} which consists of 1000 robots interacting in a limited environment and the Crazyswarm \cite{Preiss17} which consists of 50 unmanned aerial vehicles executing planned trajectories through the environment.  
Ultimately, the logistics of dealing with a large number of physical entities in a confined workspace require trade-offs: either the simplification of the environment or the use of open-loop, {\it e.g.}, pre-computed, strategies that are not adaptive or reactive to changes in the environment or internal state of the swarm.
Given these logistical challenges, experimental validation of swarming strategies  are more often conducted using small numbers of robots which do not account for large number effects \cite{Li17, Pickem17, Dorigo2013, Ipparthi17, Warkentin18}.
Unsurprisingly, experiments on small numbers of robots in controlled laboratory settings limit the types and frequency of robot-robot and robot-environment interactions. 
Furthermore, such methodologies cannot suitably evaluate the performance of the coordination strategies for arbitrarily large team sizes.  
Thus, the results may not fully account for the many factors that affect the dynamics of emergent swarming behavior.

In this work, we addressed these experimental challenges in engineered swarm systems by proposing a mixed reality experimental framework as a first significant step towards full experimental validation.
Mixed reality is the use of both virtual robots and real robots in both the simulated and real world \cite{Honig15}, that retains critical features of physical robots while enabling scaling to larger numbers of agents, or larger workspaces, without being subject to the physical limitations of resources.
The benefits that come from mixed reality include: ability to work with large numbers of robots \cite{Edwards18} and ensuring safety in human robot interactions \cite{Quinlan10, Huang19, Rosen19}.
Experiments using mixed reality come at a lower cost due to the reduced number of robots needed, while still introducing complex dynamics of the real world from a few real robots. 
Mixed reality is a significant first step towards full scale experimental validation of theoretical findings. Furthermore, the mixed reality framework provides opportunities to gain additional insights into the theory and improving experiment design.

Our current research uses mixed reality as a way to further study the controller proposed in Szwaykowska et al. \cite{Szwaykowska2016}, and to map out experimentally a complete bifurcation picture in terms of physical parameters.
In addition to uncovering the bifurcation structure of the swarm dynamics, we will focus on understanding transitions between behaviors and the impacts of adding collision avoidance.
The new experiments are done using two different platforms of interest: one uses an Unmanned Aerial Vehicle (UAV), and the other uses an Autonomous Surface Vehicles (ASV), both within a mixed reality framework. 
The use of two different platforms has several advantages.
First, it tests the universal bifurcation structure of delay coupled swarms across different platforms and vastly different time scales. 
Second, it allows for different numbers of robots and constraints to be tested safely during experimentation. 

The layout of the paper is as follows.
In section \ref{sec:method} we describe in detail the model used to control the swarm.
In section \ref{sec:experiments} we outline the experimental setup for both UAV and ASV experiments, and in section \ref{sec:results} we explain the results observed. 
In section \ref{sec:discussion} we discuss different phenomena that arose during experimentation, {\it{e.g.}}, platform differences, obstacles encountered, and multi-stability.
In section \ref{sec:conclusion} we conclude.  

\section{Methodology}\label{sec:method}

Consider a swarm composed of $N$ robots positioned in the plane, $r_i \in \mathbb{R}^2$, where $i \in {1 ... N}$. We begin by detailing the development of our single agent and swarm ensemble models.

\subsection{Single agent model}
The dynamics for each agent in the system consist of a self propulsion term, an attraction term, and a repulsion term. This can be mathematically represented as follows using the original equation proposed by Mier-y-Teran-Romero \cite{Romero2012} for the $i^{th}$ agent:

\begin{equation}
    \dot{r_i} = v_i,
\end{equation}

\noindent and

\begin{align}
\ddot{r_i} = (1 - ||\dot{r_i}||^2)\dot{r_i} - \sum_{j\in N} \nabla_{r_i} U[r_i(t),r^{\tau}_j(t)],
\label{eq:original}
\end{align}

\noindent To model the communication topology between agents, we consider
a fully connected graph model, $G = (\mathcal{E}, \mathcal{V})$, where $\mathcal{E}, \mathcal{V}$ are the set of edges and vertices, or nodes, respectively.
  We improve upon the validity of this model by having the robots communicate a delayed position $r^{\tau}_j(t) = r_j(t - \tau)$.
  Note that the robots interact with one another with a fixed time delay, $\tau$, which captures realistic finite-time effects of robot to robot communication.

We assume a harmonic interaction potential defines the attraction term

\begin{align}
  U(r_i, r_j^{\tau}) = f(r_i, r_j) + \frac{a}{2N}(r_i - r_j^\tau)^2,
  \label{harmonic_equation}
\end{align}

\noindent where $a$ is a constant, and $f(r_i, r_j)$ is a repulsion term.

In previous theoretical work the repulsion force was added to only a fraction of the agents in the experiments \cite{Szwaykowska2016}.
However, for real systems to interact safely in the world repulsion forces for all agent interactions are necessary, making it important to extend this work to consider the addition of repulsion to all agents in the swarm.
As long as the repulsion force selected is an anti-symmetric function in the neighboring robot's states, the analysis performed in section \ref{sec:ensemble_swarm_model} of the global swarm behavior will be preserved.

For the experimentation done in this paper, two anti-symmetric functions were selected. 
The original repulsion force presented in Szwaykowska et al.\cite{Szwaykowska2016} is:

\begin{align}
  f(r_i, r_j) =  c_re^{\frac{||r_i - r_j||}{l_r}},
  \label{eq:exponential_rep}
\end{align}

\noindent where $c_r$ is the strength of the repulsion, and $l_r$ is the radius of repulsion considered between agents.

However, Equation \ref{eq:exponential_rep} does not account for limitations of physical platforms, {\it{e.g.}}, max speeds or acceleration capacity. As such a sigmoid repulsion function is used:

\begin{align}
  f(r_i, r_j) = \Big(c_{r} - \frac{c_{r}}{1+e^{-k(|r_i - r_j|-R_{rep})}}\Big)\frac{r_i - r_j}{|r_i - r_j|}.
  \label{eq:sigmoid_repulsion}
\end{align}

\noindent where $c_{r}$ is the maximum repulsion strength, $R_{rep}$ is the inter agent distance at which the repulsion force is at half strength, and $k$ represents how quickly the magnitude of the repulsion force switches from maximum strength to zero.
Note that the repulsion term is independent of the delay since the interactions for repulsion are local in space. 
 
 \subsection{Ensemble Swarm Model}
 \label{sec:ensemble_swarm_model}
The mean field of a swarm is computed by taking $R=\frac{1}{N}\sum_{i=1}^{N}r_i$ to denote the center of mass, and consider the limit as $N \rightarrow \infty$.
From a mean field analysis of Equation \ref{eq:original}, analytical expressions can be derived for different swarm states\cite{Romero2012}.
The swarm state is the global representation of the entire swarm evaluated by observing the center of mass of the swarm, $R$, in lieu of the position of all agents.
The state of the robot, $r_i$, is the individual dynamics for the local behavior of the robot. 

The mean field of the  original controller for the swarm behavior was shown theoretically to have several bifurcating regions in parameter space.
Each region implies the stability of certain swarm behaviors including: flocking, ring, and rotating swarm states.

\begin{figure}
  \centering
    \subfigure[Bifurcation Diagram]{
      \label{fig:dim_bifurcation}
      \includegraphics[width=\linewidth]{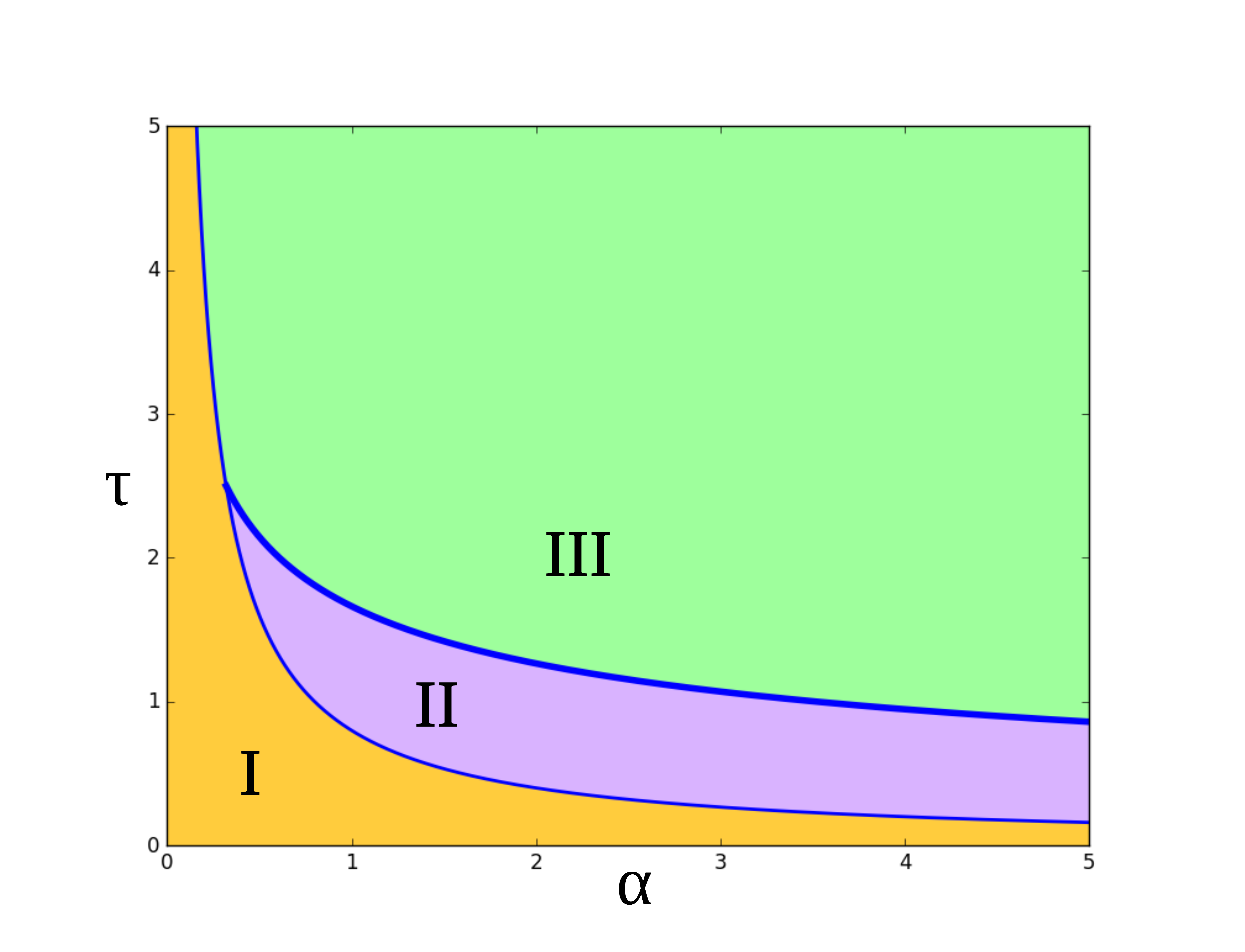}
    }
    \subfigure[Swarm States]{
      \label{fig:sim_example}
      \includegraphics[width=0.8\linewidth]{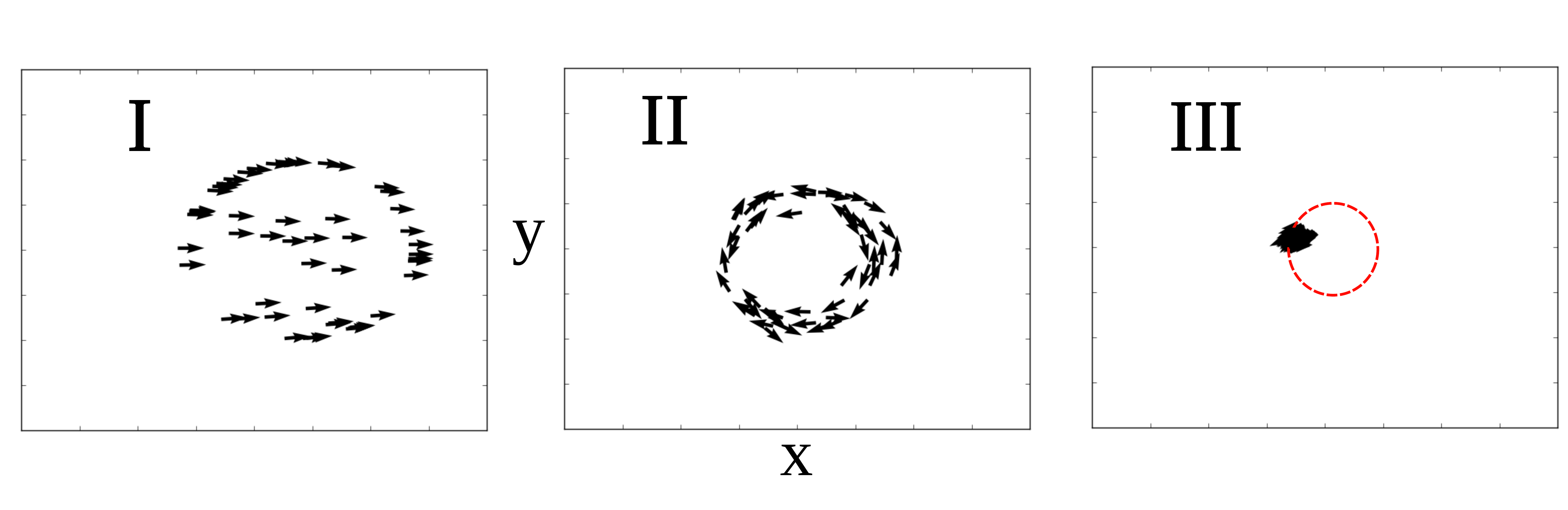}
    }
  \caption{
    Figure \ref{fig:dim_bifurcation}: A bifurcation diagram as a function of communication delay, $\tau$\,(s), and coupling amplitude, $\alpha$\,(1/s$^2$) for the parameters used in UAV experiments: $\beta = 20.0$\,s/m$^2$, $v_g = 0.2$\,m/s.
    The solid curves are predicted from the mean field equations.
    The swarm is in a translating state with parameters from region I, and the swarm is in the rotating state with parameters from region III.
    In region II, the swarm is in a ring state, which also appears for parameters in region I and III.
    Note that for the ASV experiments different parameters were used, as such, a different bifurcation graph occurred meaning transitions between swarm states occurred for different parameter combinations.
    The ASV parameters were: $\beta = 18.0$\,s/m$^2$ and $v_g = 0.047$\,m/s.
    Additionally, repulsive forces impact the bifurcation structure and transition points, resulting in discrepancies from the mean-field predictions.
    Figure \ref{fig:sim_example}: The three possible swarm states are shown using 50 simulated agents (I Translating, II Ring, III Rotating).
}
    \label{fig:dimensional_bifurcation_figure}
\end{figure}

 Figure \ref{fig:dim_bifurcation} is the converted dimensional version of the original dimensionaless bifurcation structure proposed in Szwaykowska et al \cite{Szwaykowska2016}.
  The transition between region II and III was theoretically predicted by a Hopf bifurcation curve of the mean field, and a pitchfork bifurcation curve is predicted to separate regions I and II.
Figure \ref{fig:sim_example} illustrates the three basic modal patterns of the swarm behavior as a function of the coupling strength, $\alpha$, and communication delay, $\tau$.

For the application of the dynamical model to the real world, we consider the dimensionalized equation for each agent as follows:

\begin{align}
  \ddot{r_i} &= \beta(v_g^2 - ||\dot{r_i}||^2)\dot{r}_i - \frac{\alpha}{N}\sum_{j=1,j \neq i}^N(r_i - r_j^\tau) + \sum_{j=1, j\neq i}^N\nabla_r f(r_i, r_j),
  \label{eq:dimensional}
\end{align}

\noindent where $v_g$\,m/s is the asymptotic velocity of the agent in the absence of coupling, $\alpha$\,1/s$^2$ is the coupling strength, $\beta$\,s/m$^2$ is a dimensional factor, and $\nabla_{r_i} f(r_i, r_j)$ is the repulsion force.

In order to further study the theoretically predicted swarm states, it was necessary to compute a new dimensional bifurcation diagram, which considers the physical parameters used in Equation \ref{eq:dimensional}. 
To achieve Figure \ref{fig:dim_bifurcation} the following conversions were used: 

\begin{align}
  t' = \beta v_g^2 t, \\
  r'_i = \beta v_g r_i, \\
  a = \frac{\alpha}{\beta^2v_g^4},
  \label{eq:parameter}
\end{align}

\noindent where $a, t',$ and $r'_i$ are dimensionless.

The conversion of the bifurcation diagram allows for regions of interest to be isolated, specifically around the regions of uncertainty along the Hopf bifurcation.
The original mean-field analysis does not change through conversion, which is demonstrated in the appendix where the mean field was re-derived for Equation \ref{eq:dimensional}.

The three desired swarm states are highlighted in Figure \ref{fig:dim_bifurcation}.
In region I the swarm is in a translating state, where all agents are in alignment going in one direction.
In region II the swarm is in a ring state, where all agents move about a stationary center of mass, and in region III the swarm  is in a rotating state, where all agents cluster and move as a collective on a circular orbit around the origin. 
Along the boundaries of each region the swarm state will transition from one swarm state to another. This is referred to as the transition between swarm states. 
The possibility of multi-stable co-existing rotational patterns along these regions makes experimental verification important to further understand the limitations of the mean field analysis. 

\section{Experiments}\label{sec:experiments}

\begin{figure}
  \centering
  \subfigure[Ascending Technologies Inc. Pelican Quadrotor]{
    \label{fig:UAV}
    \includegraphics[width=0.9\linewidth]{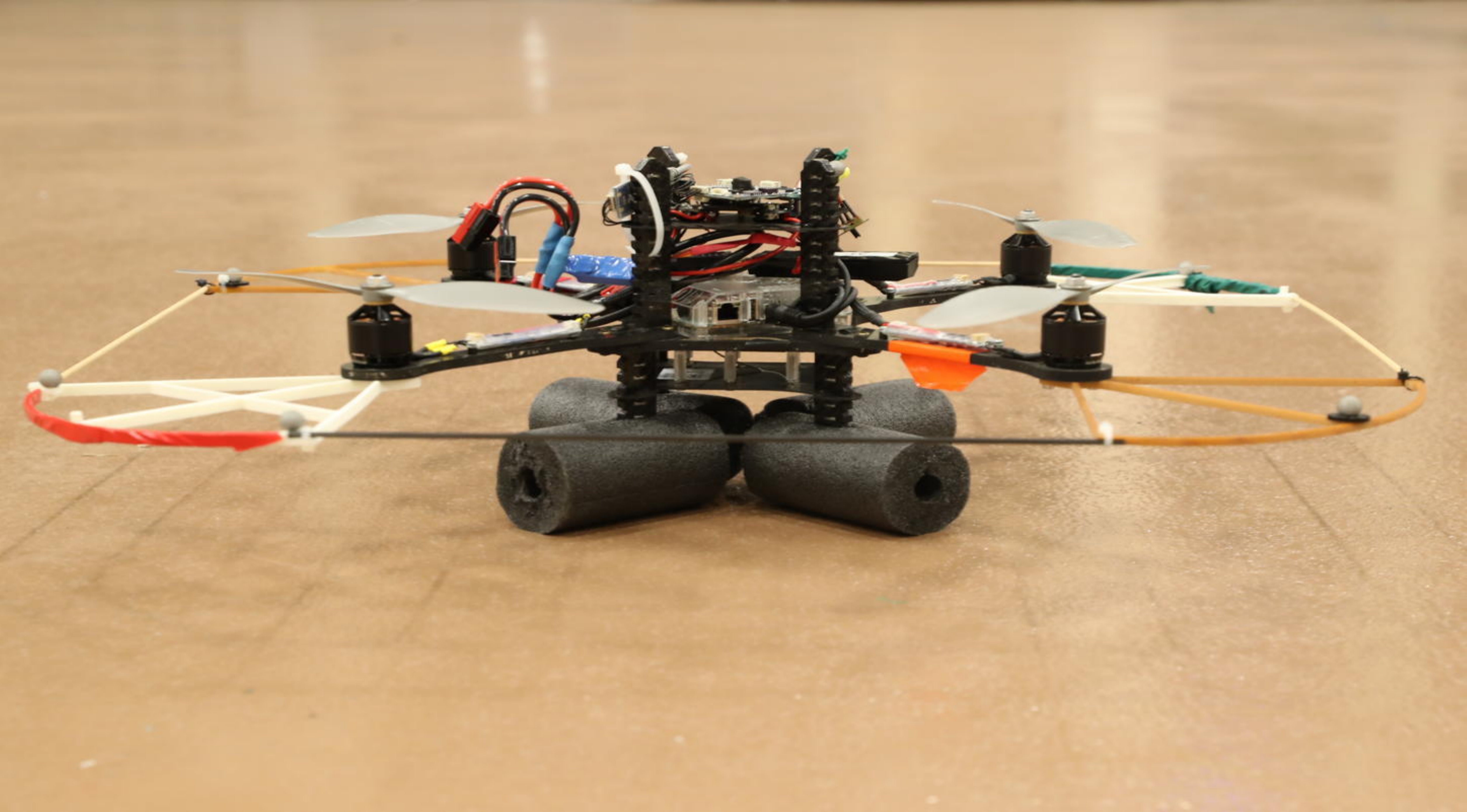}
  }
  \subfigure[Autonomous Surface Vehicle]{
      \label{fig:ASV}
      \includegraphics[width=0.9\linewidth]{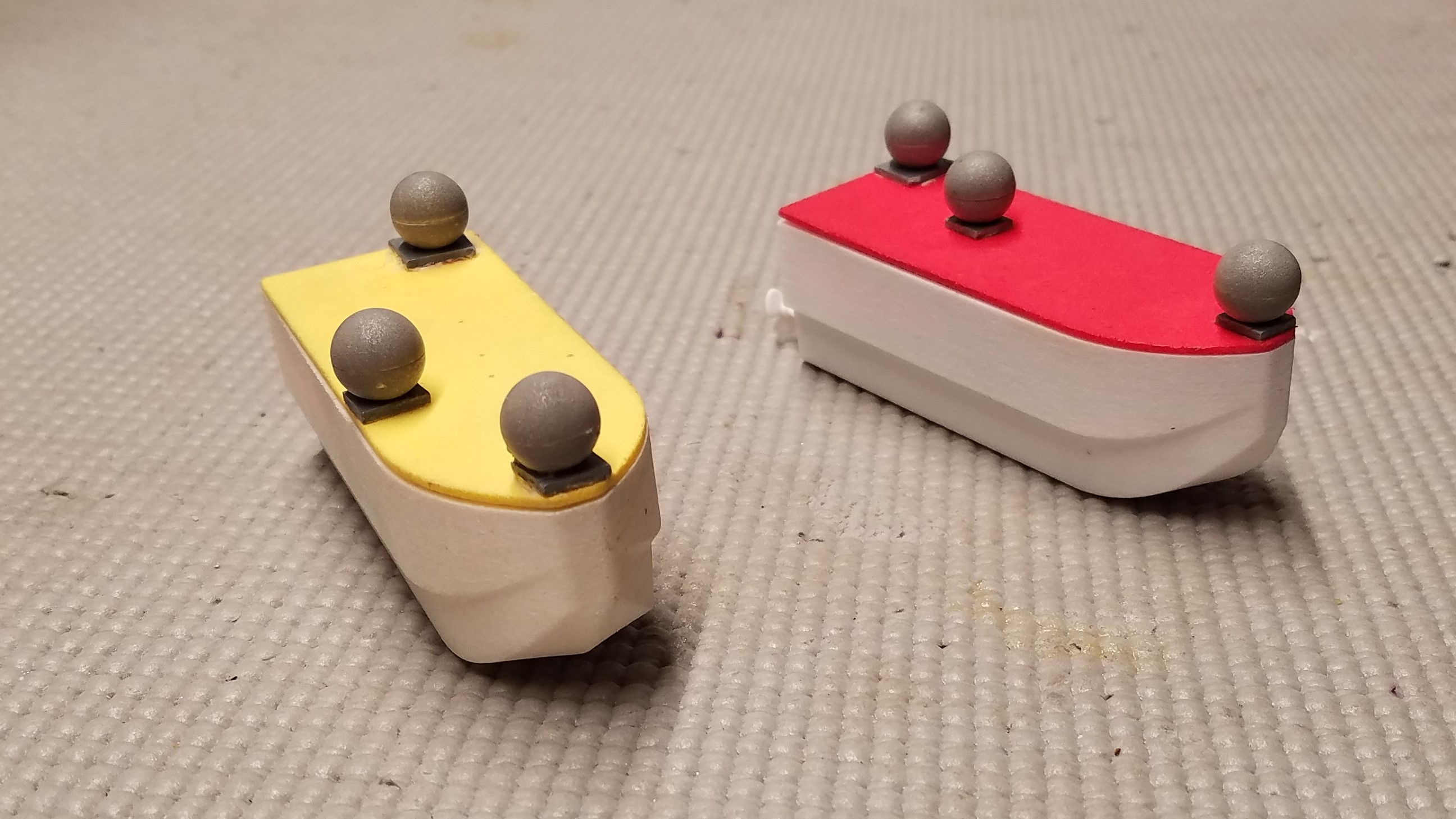}
  }
  \caption{
    Figure \ref{fig:UAV} The Ascending Technologies Inc. Pelican Quadrotor, AscTec Pelican quadrotor.
    Figure \ref{fig:ASV} Autonomous Surface Vehicle, ASV, built at the University of Pennsylvania. 
  }
\end{figure}

The objective of our experiments is to test the mean field predictions described in Section \ref{sec:method} in swarms composed of a few real robots.
  Discrepancies can then be used to build more accurate theories and analysis for future experiments with larger numbers of real robots.
Mixed reality experiments were conducted using both the Ascending Technologies Inc. Pelican quadrotor shown in Figure \ref{fig:UAV} and custom built ASVs shown in Figure \ref{fig:ASV}.
We show experimentally all three swarm states (translating, ring, and rotating), along with the transition between swarm states as predicted by the bifurcation diagram in Figure \ref{fig:dim_bifurcation} within the mixed-reality framework.
We describe the details of our mixed reality architecture, experimental platforms, and experimental methodology in the following sections.

\begin{figure}
  \centering
  \subfigure[Mixed reality Setup]{
    \label{fig:mixed_reality}
    \includegraphics[width=\linewidth]{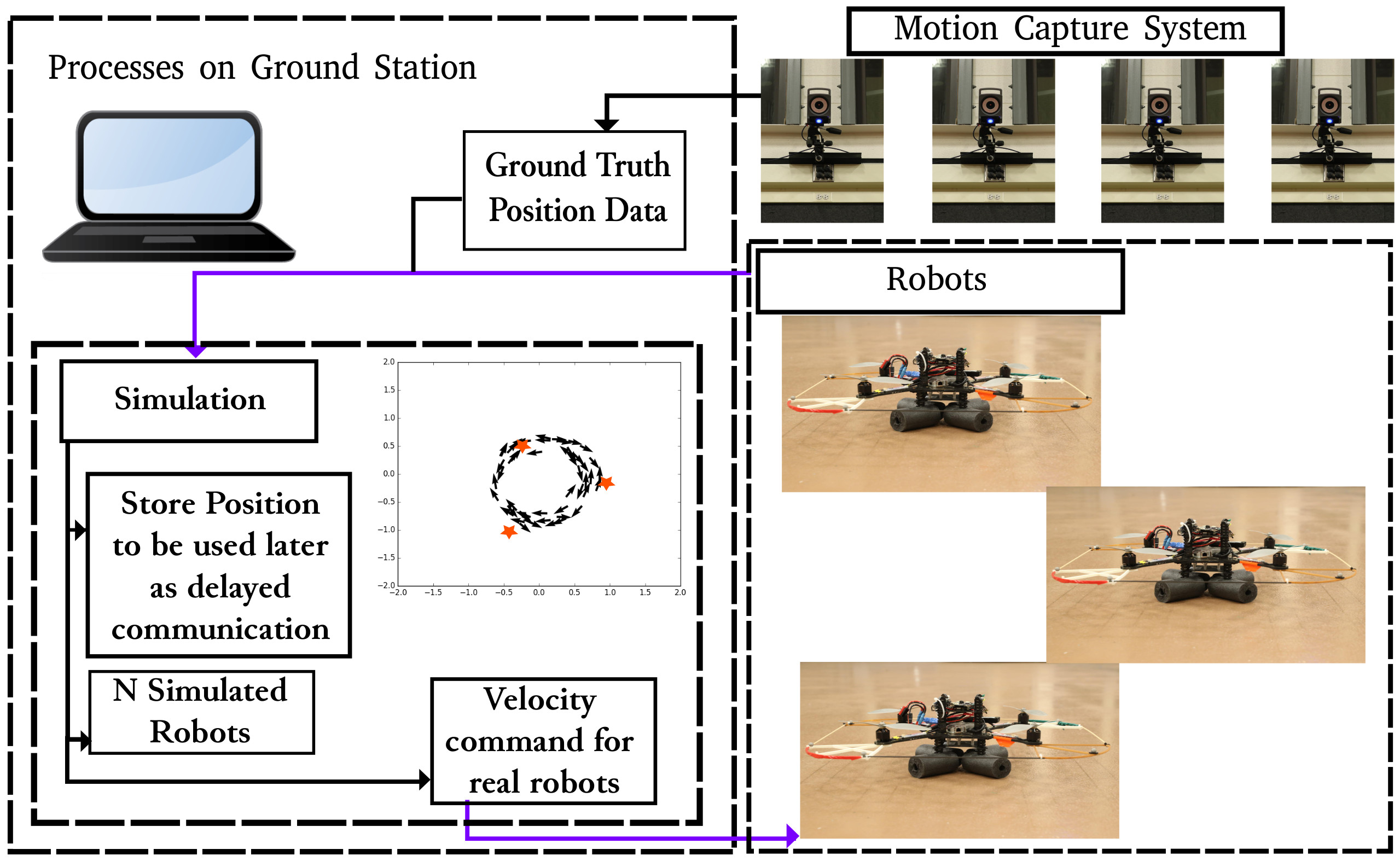}
  }
  \subfigure[Physical Example]{
    \label{fig:mr_robot_example}
    \includegraphics[width=\linewidth]{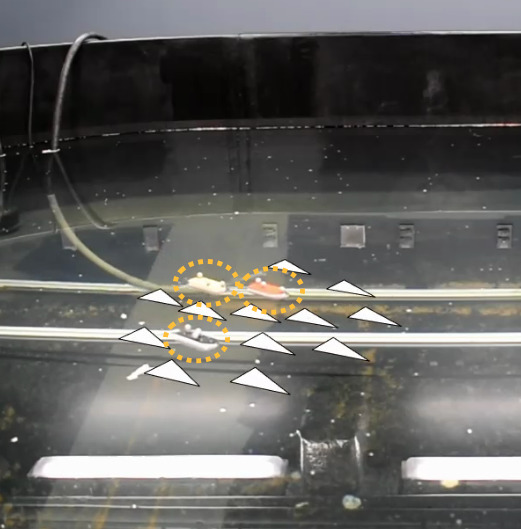}
  }
  \caption{
    Figure \ref{fig:mixed_reality}: An outline for the mixed reality platform, where a ground station computer maintains the simulation and the state of the robot. 
    The simulator is informed about the robot's state using a motion capture system to estimate of the robot's position, and the simulator sends new control outputs to the UAV. 
    Figure \ref{fig:mr_robot_example}: An example mixed reality setup for 3 ASVs and 12 simulated robots projected into image space.
  }
\end{figure}

\subsection{Mixed reality system architecture} 
An outline of a mixed reality system is in Figure \ref{fig:mixed_reality}, and an example of a mixed reality experiment is depicted in Figure \ref{fig:mr_robot_example}.
In building a mixed reality, global positions of the real robots are necessary, {\it e.g.}, GPS, infrared camera system, SLAM.  
The ground truth position of the robots are provided to the simulator, which maintains all simulated agents. 
The resulting response outputs for the robots are computed and sent to the robot from the simulator, considering interactions with all simulated agents. 

The mixed reality system is controlled by the simulator, which maintains the positions of the simulated robots and updates the positions of the real robots based on ground truth information, as seen in Figure \ref{fig:mixed_reality}. 
The simulator workspace is defined as an unbounded region with no obstacles.
The origin of the simulated workspace corresponds to the origin of the physical workspace. 
Delayed information is stored in a fixed length list for each agent which holds previous positions.
  The length of the list corresponds to the amount of delay specified: larger $\tau$ is a longer list, and smaller $\tau$ is a shorter list. 
The last entry in the list represents the delayed information received by an agent.
All agents leveraged the global knowledge of the simulator to compute the distances between agents.
Sensing between agents was abstracted away, which allowed for the focus of the results to be on the swarm states.
From the simulator, new control commands were sent to the UAV/ASVs based on delayed information of simulated agents, as if they were in the world.
Mixed reality allowed the UAV/ASVs to express the swarm behavior without the risk of multiple real robots interacting in unpredictable ways or the cost of running a multi-robot experiment. 

\subsection{Experimental Platforms}
In this work we employed two experimental platforms: unmanned aerial vehicles (UAVs) and autonomous surface vehicles (ASVs).
The validation using two separate platforms shows the applicability to any vehicle platforms whose dynamics can be abstracted into Equation \ref{eq:original}.
The UAV, Figure \ref{fig:UAV}, is a quadrotor vehicle, which is equipped with an Odriod for onboard computing, WiFi communication, and an inertial guidance system (AscTec Autopilot).
The vehicle is approximately $65.1$\,cm $\times$ $65.1$\,cm and $18$\,cm tall and weighs $1.65$kg.
The workspace is a $15$m $\times$ $10$m $\times$ $8$m room.
The ASVs, Figure \ref{fig:ASV}, are differential drive surface vehicles equipped with a micro-controller board, XBee radio module, and an inertial measurement unit (IMU).
The vehicles are approximately $12$ cm long and have a mass of about $45$g each.  The ASVs are deployed in a $3$ m $\times$ $5$ m $\times$ $1$m oval tank.
Both workspaces include an infrared (IR) camera system to provide robot localization information. 

\subsection{Experimental Methodology}
The experiments with the UAVs consisted of 1 physical and 49 simulated robots.
The motions of simulated agents are updated using a double-point-integrator with Equation \ref{eq:dimensional}.
For the translating swarm state, and the transition swarm state experiments the original configurations consisted of all agents facing in the same direction in a fixed pattern, with the same initial input velocity of 0.2\,m/s. 
For the ring and rotating swarm state experiments, the UAV was placed at $[0, 0, 0]$ with the simulated robots placed around a rough ring shape with initial velocities in x and y selected between $[-0.3, 0.3]$\,m/s.
To achieve transition between swarm states $\Delta \tau$ was added every 10\,s to $\tau$.
Details of experimental parameters are in Table \ref{table:UAV_params}.
Parameters were chosen for Equations (\ref{eq:parameter}) to satisfy maximum speed constraints for the real robots and finite workspace size. 
  Using the theoretical equations predicting the ring and rotating state radii, parameters were selected which provided the desired radii size ($\approx 0.75$\,m).
  The parameters were tested in simulation before being tested in experiments.

  A small amount of repulsion was introduced to the experiment, Equation \ref{eq:exponential_rep}, where $c_r = 1.2$ and $l_r = 0.01$.
  To accommodate the low repulsion forces, the experiments were constrained to a two dimensional slice such that each real and simulated agent were on a unique plane.
  The interactions between agents were achieved by projecting all agents onto the same two dimensional plane, and velocities for the robot were a two dimensional velocity with an additional altitude component \cite{Mellinger11}. 

\begin{table}
\begin{tabular}{|c|c|c|c|c|c|c|c|}
     \hline
     Swarm State & Duration & $\alpha$\, 1/s$^2$ & $\tau$\,s & $\beta$\,s/m$^2$ & $v_g$\,m/s & $\Delta \tau$ \\
    \hline
     Translating & 30 & 0.01 & 0.01 & 20.0 & 0.2 & 0.0 \\ 
    \hline
     Ring & 90 & 0.09 & 2.5 & 20.0 & 0.2 & 0.0 \\ 
    \hline
     Rotating & 90 & 1.0 & 4.0 & 20.0 & 0.2 & 0.0 \\ 
    \hline
    Tran - Ring - Rot & 200 & 1.5 & 0.01 & 20.0 & 0.2 & 0.3 \\
    \hline
\end{tabular}
\caption{Experimental parameters for UAV mixed reality experiments}
\label{table:UAV_params}
\end{table}

Next we considered the impacts of adding more robots to the swarm, thus requiring stronger repulsive forces. 
To do this we used a team of ASVs shown in Figure \ref{fig:ASV}.
The experiments consisted of 3 physical and 12 simulated robots. 
These experiments were specifically designed to observe transitions between swarm states by modifying the delay provided to the system. 
These were untested properties of the swarm model, and mixed reality allowed for testing theoretical predictions.

The parameters for the ASV mixed reality experiments are in Table \ref{table:ASV_params}. 
For all experiments starting in the translating swarm state, the ASVs were initialized in a formation based on relative positions and moving forward at the desired speed $v_g$.
For the experiment starting in the ring swarm state the ASVs were initialized at three equidistant points on a circle of radius 0.3\,m pointing counter clockwise and tangent to the circle, and the virtual agents were initialized with random heading at points along the circumference of the circle with a normally distributed amount of noise added to their positions.
Due to the physical constraints of the ASVs, sigmoid repulsion\cite{repulsion_details} (Equation \ref{eq:sigmoid_repulsion}) was used in Equation \ref{eq:dimensional}.

\begin{table}
\begin{tabular}{|c|c|c|c|c|c|c|c|}
    \hline
     Swarm State & Duration & $\alpha$\, 1/s$^2$ & $\tau_0$\,s & $\beta$\,s/m$^2$ & $v_g$\,m/s & $\tau_1$\,s \\
    \hline
     Tran-Ring-Rot &  660 & 0.01 & 0.0 & 18.0 & 0.0471  & 10, 35 \\ 
    \hline
     Tran-Rot-Tran & 400 & 0.01 & 0.0 & 18.0 & 0.0471 &  35 \\ 
    \hline
     Tran-Ring & 300 & 0.01 & 0.0 & 18.0 & 0.0471 & 10 \\ 
    \hline
     Ring-Rot & 380 & 0.01 & 10.0 & 18.0 & 0.0471 & 35\\
    \hline
\end{tabular}
\caption{Experimental parameters for ASV mixed reality experiments}
\label{table:ASV_params}
\end{table}

\subsection{Experimental Evaluation}
Evaluation of all experiments was done using swarm polarization, which is a measure of alignment between agents in a swarm.
Swarm polarization is computed as follows:
\begin{align}
  sp = \frac{||\sum_i r_i||}{\sum_i ||r_i||}.
\end{align}
This metric evaluates the swarm state, for example; in the ring state all the individual positions of the agents will cancel out resulting in a swarm polarization of $\approx 0$, while in the rotating and translating states the swarm is aligned resulting in a swarm polarization of $\approx 1$.  
We note that swarm polarization has been proposed in earlier works, \cite{Armbruster16, Tunstrom13}, and is a measure of alignment, but can be computed in different ways.

In addition to swarm polarization, the velocity of the center of mass and the acceleration of the center of mass were computed.
These metrics aided in determining the difference between the translating and rotating swarm states, which have the same swarm polarization value.
The rotating swarm state has a high velocity and acceleration of the center of mass, while in the translating swarm state the center of mass velocity is high but the acceleration is low. 
Velocity of the center of mass is computed as follows:
\begin{align}
  c_{vel} = ||\frac{\sum_i^N v_i}{N}||,
\end{align}

\noindent and the acceleration of the center of mass is:

\begin{align}
  c_{acc} = ||\frac{\sum_i^N \dot{v_i}}{N}||.
  \end{align}

Additional comparisons were done using theoretically predicted results from equations in Szwaykowska et al. \cite{Szwaykowska2016}, and experimental results for the ring swarm state radius and period along with the rotating swarm state radius and period are presented in Figure \ref{theory_v_experiment_figure}.
These equations are listed in the appendix. 

\section{Results}\label{sec:results}
\subsection{UAV Simulation of Parameters in Different Bifurcation Regions}
\begin{figure}
  \centering
  \includegraphics[width=\linewidth]{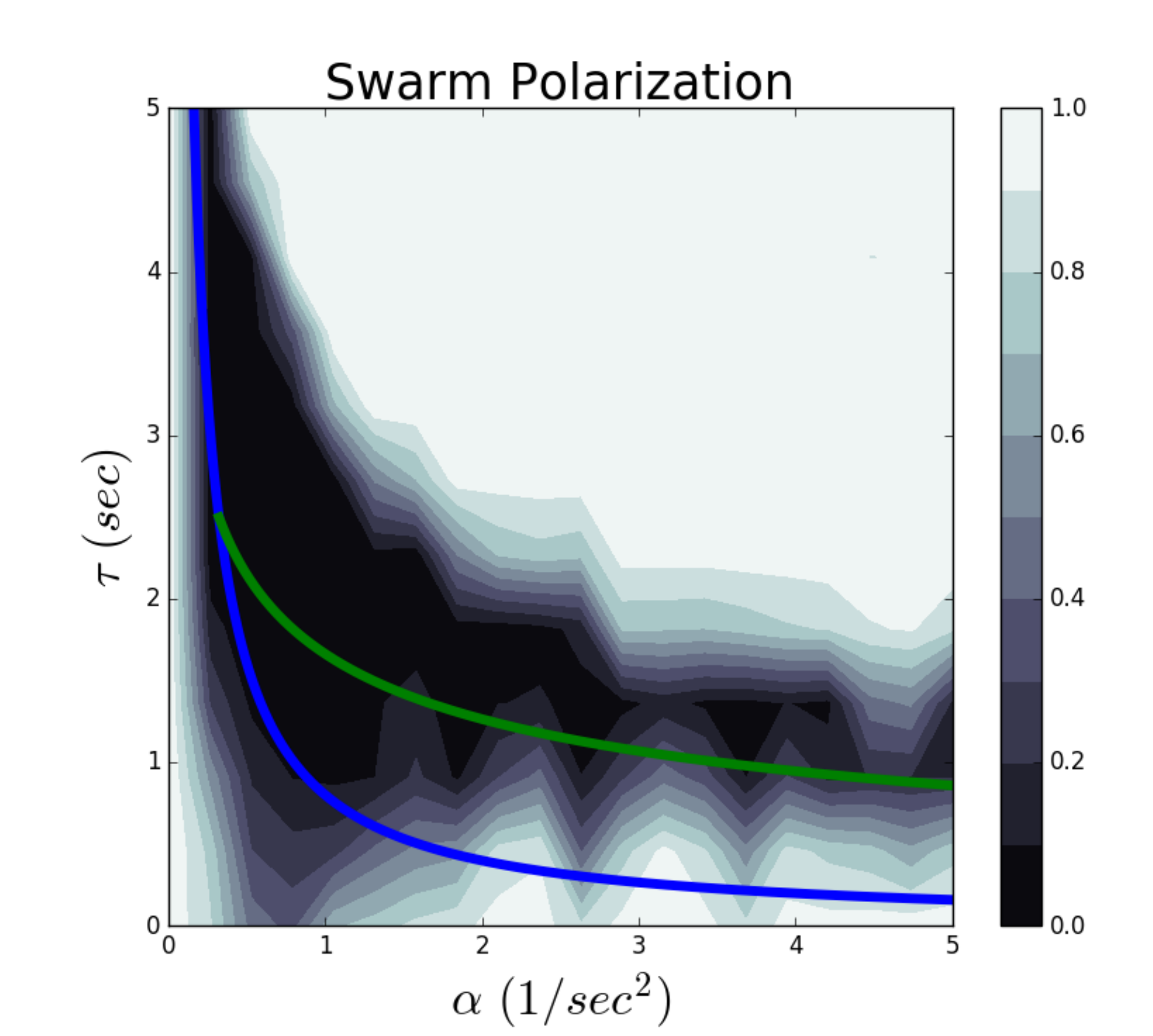}
  \caption{
    The resulting swarm polarization from 240 simulation trials with 50 simulated agents.
    Each trial was 100\,s and used UAV parameters, $\alpha \in [ 0.0, 4.75]$\,1/s$^2$, $\tau \in [0.01, 5.51]$\,s, $\beta = 20.0$\,s/m$^2$, and $v_g = 0.2$\,m/s.
    For each $\alpha$, $\tau$ was updated by 0.5\,s.
    At the end of the range for $\tau$, $\alpha$ was updated by 0.25\,1/s$^2$.
    Color is a representation of the final swarm polarization at the end of an experiment.
    The trends of the contours while slightly off of the theoretically predicted bifurcation curve still meet the general trend of the plot. A transition from white to black to white in the plot, corresponds to the transition from the translating to ring to rotating swarm states. 
  }
  \label{bifurcation_data_figure}
\end{figure}

Simulation trials using UAV parameters were executed with different combinations of $\alpha$ and $\tau$, testing the theoretically predicted bifurcation regions in Figure \ref{fig:dim_bifurcation}.
The simulation trials used parameters corresponding to the UAV experiments. 
The trials ran for 100\,s, with $\beta = 20.0$\,s/m$^2$ and $v_g = 0.2$\,m/s.
The initial $\alpha = 0.0$\,1/s$^2$ and $\tau = 0.01$ were the parameters used for the first trial.
At the completion of a trial $\tau$ was updated by $0.5$.
The range of $\tau$ was $\tau \in [0.01, 5.51]$\,s for each value of alpha.
When a trial for each value in the range of $\tau$ was complete $\alpha$ was updated by 0.25. 
The range of $\alpha$ was $\alpha \in [0.0, 4.75]$\,1/s$^2$.  
In total, two hundred and forty simulation trials were run with 50 simulated agents.
Figure \ref{bifurcation_data_figure} shows the resulting plot of the final swarm polarization at the end of each simulation, for each set of parameters,  which shows fair qualitative agreement with mean field predictions.

Regions of multi-stability exist along the transition between swarm states, meaning that for certain initial conditions the swarm will transition to another swarm state or it may stay in the same swarm state \cite{Romero2014}. 
This is clearly seen by the jagged edge along the bifurcation in Figure \ref{bifurcation_data_figure}.
This result showed that more experimentation using physical robots needed to be done, exploring each of the three swarm states along with transitions between the swarm patterns.

\subsection{UAV Mixed Reality Results}\label{sec:uav_mr}
The first set of experimental results used a UAV mixed reality framework. 
Experiments of the swarm in the translating swarm state were achieved using parameters from region I in Figure \ref{fig:dim_bifurcation}, where all agents move with the same average velocity.
The swarm would quickly start moving out of the translating swarm state and into the ring swarm state in the last seconds of the experiment. 
This was due to the instability of the swarm when translating, from the introduction of any noise in the system, which can switch the system out of the translating swarm state and into the ring swarm state.

\begin{figure*}
  \subfigure[ Ring Swarm State Period]{
  \includegraphics[width=0.4\linewidth]{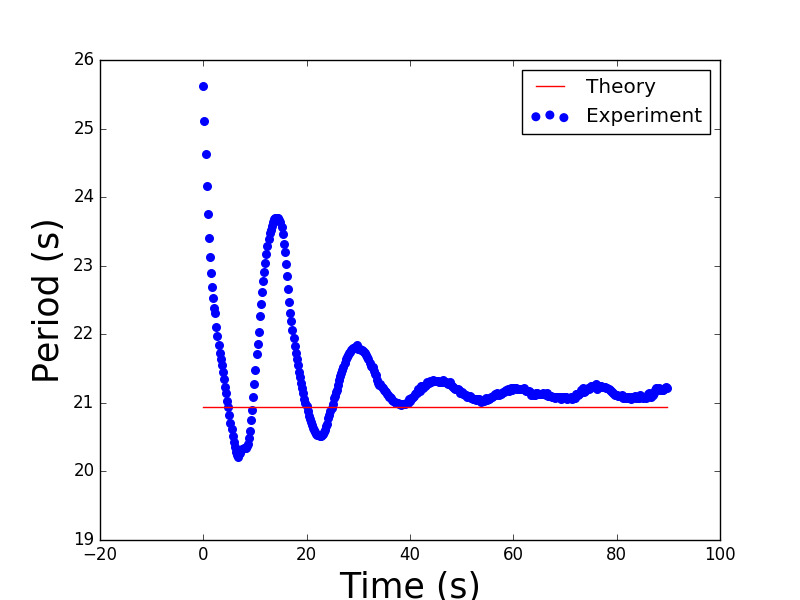}
    \label{fig:ring_period}
  }
  \subfigure[ Ring Swarm State Radius]{
  \includegraphics[width=0.4\linewidth]{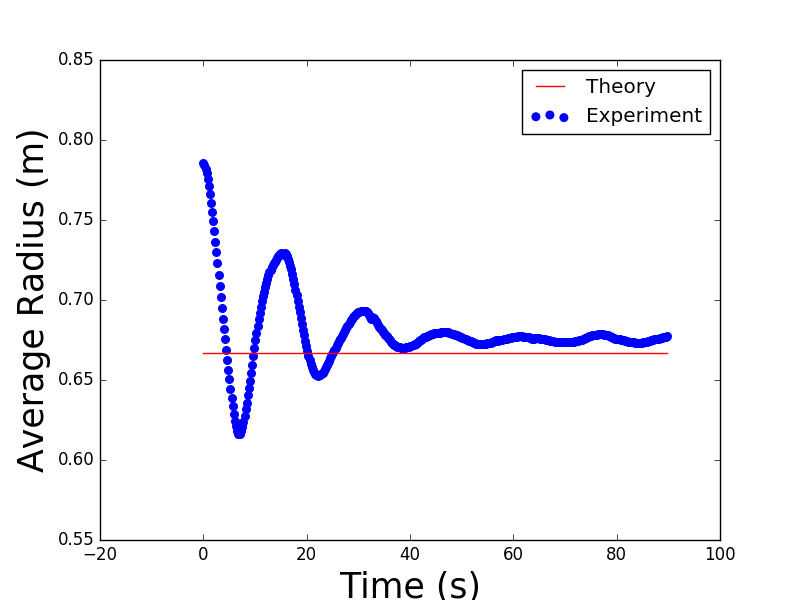}
    \label{fig:ring_radius}
  }
  \subfigure[ Rotating Swarm State Period]{
  \includegraphics[width=0.4\linewidth]{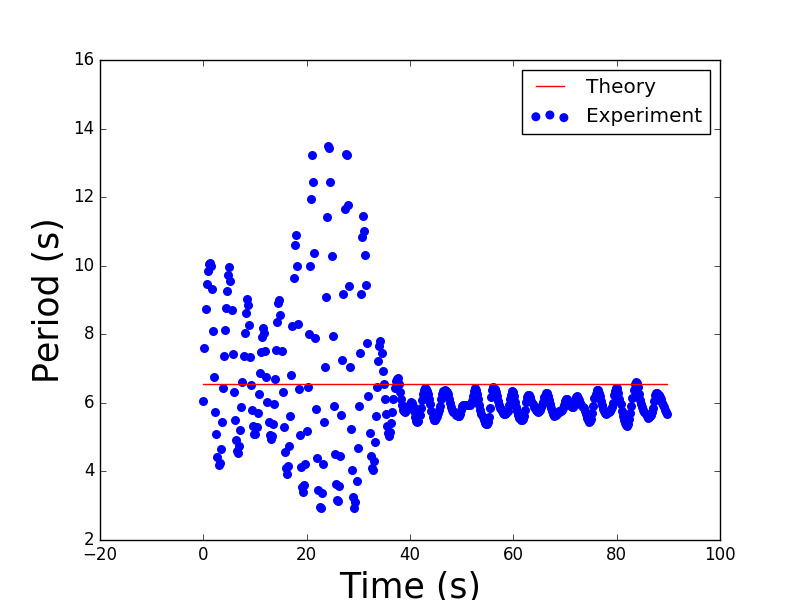}
    \label{fig:rot_period}
  }
  \subfigure[ Rotating Swarm State Radius]{
  \includegraphics[width=0.4\linewidth]{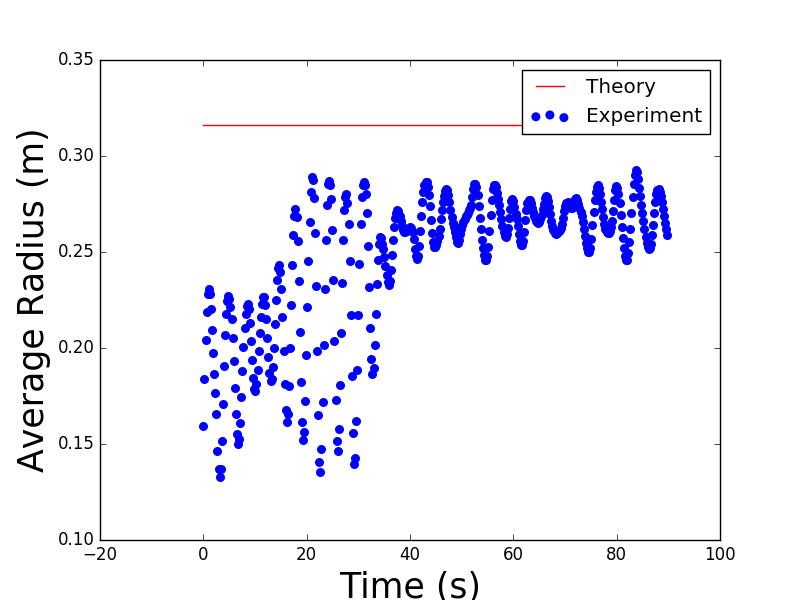}
    \label{fig:rot_radius}
  }
  \caption{
    Figure \ref{fig:ring_period} and \ref{fig:ring_radius}: Results from a 90.0\,s ring experiment, which compared the theoretically predicted radius and period (red line), to the average for the swarm of the exhibited radius and period during a ring experiment (blue points).
    Figure \ref{fig:rot_period} and \ref{fig:rot_radius}: Results from a 90\,s rotating experiment, which compared the theoretically predicted radius and period (red line), to the average for the swarm of the exhibited radius and period during a rotating experiment (blue points).}
  \label{theory_v_experiment_figure}
\end{figure*}

Experiments where the ring swarm state was expressed used parameters from region II in Figure \ref{fig:dim_bifurcation}, where all agents move, clockwise or counter-clockwise, around a stationary center of mass.
To further test that the dynamic model matched experimental results, comparisons were made between the theoretically predicted radius and the period of the ring swarm state to the experimental values. 
Figures \ref{fig:ring_period} and \ref{fig:ring_radius} show the average swarm ring period and radius experimental result (blue points) and the theoretically predicted value (red line).
These experiments show that at steady-state the average swarm ring radius converges to approximately 0.68\,m and the average swarm ring period converged to approximately 21.22\,s, these values were both close to the theoretically predicted values of 0.66\,m and 20.94\,s. 
To achieve the experimental ring radius and ring period, parameters were selected which put the swarm clearly in the ring swarm state.
The theoretical values were computed based on the parameters selected for the experiments.
The proximity of the measured ring radius and ring period comes from ensuring well calibrated experiments which expressed the desired behavior.  
This result qualitatively supports the comparison between the swarm theory and the experimental results.

The rotating swarm state experiments used parameters from Region III in Figure \ref{fig:dim_bifurcation}.
During the rotating swarm state experiment all agents clustered together and moved in a collective around a stationary point.
Figures \ref{fig:rot_period} and \ref{fig:rot_radius} show the average radius and period of the swarm in the rotating state, comparing the theoretically predicted values (red line) to the experimental results (blue points).
The theoretical period was 6.55\,s and the theoretical radius was 0.32\,m, compared with the converged average of the experimental period was 5.68\,s and the experimental radius was 0.26\,m. 
Similarly to the experiments with the swarm in the ring state, parameters were specifically selected to put the swarm clearly in the rotating swarm state. 
The measured rotating radius and rotating period are close to theoretical predictions because of well tuned experiments resulting in the desired behavior. 
The plots highlight that the swarm does converge to the theoretically predicted swarm behavior even with the addition of a real robot.

\begin{figure}
  \centering
  \includegraphics[width=\linewidth]{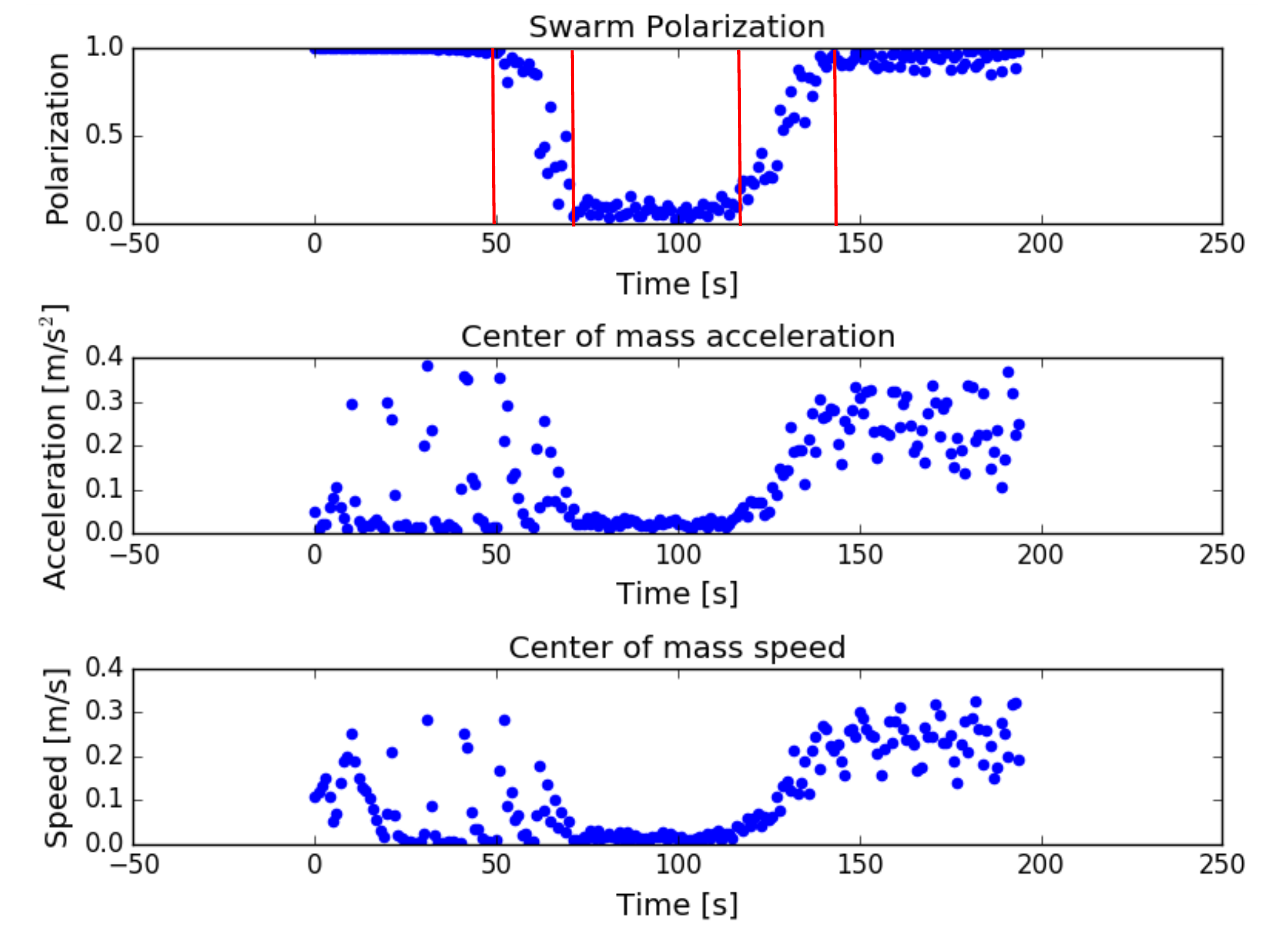}
  \caption{
    Swarm polarization, center of mass acceleration, center of mass speed for an experiment with 1 UAV and 49 simulated robots. A 200\,s experiment transitioning between translating, ring, and rotating swarm state, where the experiment started with the following parameters $\alpha = 1.5$\,1/s$^2$, $\tau_0 = 0.01$\,s, $\beta = 20.0$\,s/m$^2$, $v_g = 0.2$\,m/s, and every 10 seconds $\tau$ was updated by $\Delta \tau = 0.3$\,s.
    The increment in tau caused the transition between the translating state to the ring state to occur between $\tau = [1.81, 2.41]$\,s and the transition from the ring to the rotating state to occur at $\tau= [3.61, 4.21]$\,s, labeled on the plot by red lines.
    }
  \label{switching_exp_data_figure}
\end{figure}

Finally, Figure \ref{switching_exp_data_figure} depicts the experimental results from an experiment transitioning through all three swarm states.  
The translating swarm state had swarm polarization of $\approx 1$, a low center of mass acceleration, and a high center of mass speed.
The speed of the center of mass decreased rapidly, with the addition of greater values of $\tau$ causing the swarm to begin to switch into the ring swarm state. 
The transition to the ring swarm state occurred between $\tau = [1.8, 2.41]$\,s, where the swarm had a stationary center of mass resulting in 0 swarm polarization, 0m/s center of mass speed, and 0 m/s$^2$ center of mass acceleration.
The final transition to the rotating swarm state occurred between $\tau = [3.61, 4.21]$\,s, resulting in a swarm polarization of $\approx 1$, a high center of mass acceleration, and a high center of mass speed.
Although qualitatively accurate, the existence of small discrepancies between predicted and measured dynamics for a swarm with a single real UAV suggests the need for a more accurate description of the UAV dynamics.

\subsection{ASV Mixed Reality Results}
ASV mixed reality experiments were performed to further investigate the transition between swarm states, and safely increase the number of robots.
These experiments consisted of 3 ASVs and 12 simulated robots. 
Each experiment had multiple trials of different types of sigmoid repulsion \cite{repulsion_details}.
The first ASV mixed reality experiments tested the transitions between all three swarm states. 
Additional experiments tested the transitions from translating swarm state to rotating swarm state, from the translating swarm state to the ring swarm state, and finally from the ring swarm state to the rotating swarm state.

\begin{figure}
  \centering
  \includegraphics[width=\linewidth]{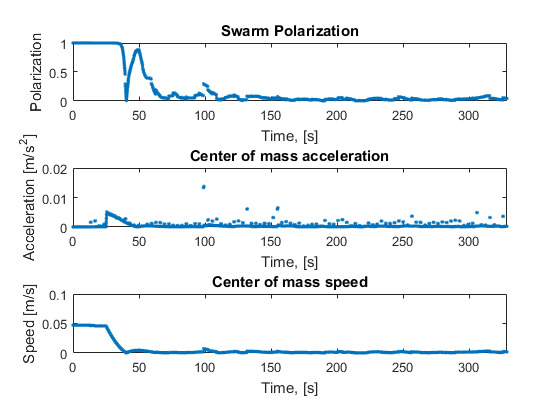}
  \caption{Swarm polarization, center of mass acceleration, and center of mass speed for the ASV swarm using local sigmoidal repulsion\cite{repulsion_details} during a mixed reality experiment.
    A delay of 10s was introduced at t = 25s causing a translating to ring transition.
    The spike in polarization at t = 50s is caused by the agents momentarily reversing direction and translating in an aligned formation before breaking up.}
  \label{upenn_exp_type2rep_swarm_stats}
\end{figure}

\begin{figure}
  \centering
  \includegraphics[width=\linewidth]{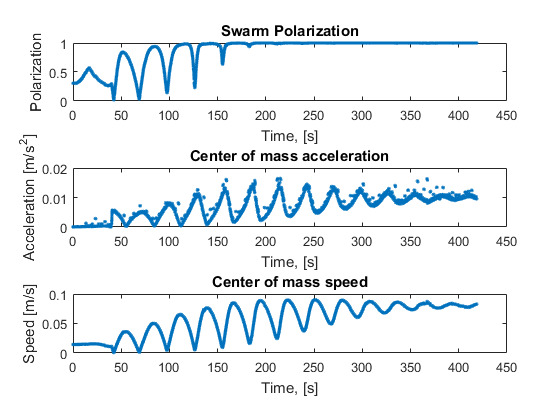}
  \caption{
    Swarm polarization, center of mass acceleration, and center of mass speed for the ASV swarm using global sigmoidal repulsion\cite{repulsion_details} during a mixed reality experiment.
    A delay of 40s was introduced at t = 60s causing a ring to rotating swarm state transition.
    The oscillation in all measures was the result of the motion of the agents moving on a gradually widening ellipse.
    When the eccentricity of the ellipse was high, the agents slowed down as they changed direction (not necessarily the same direction) at the ends of the ellipse, and sped up near the center of the ellipse.
    This behavior gradually smoothed out as the motion approached that of a circle for large t.}
  \label{upenn_exp_type1rep_swarm_stats}
\end{figure}

\begin{figure} 
 \centering
 \includegraphics[width=0.8\linewidth]{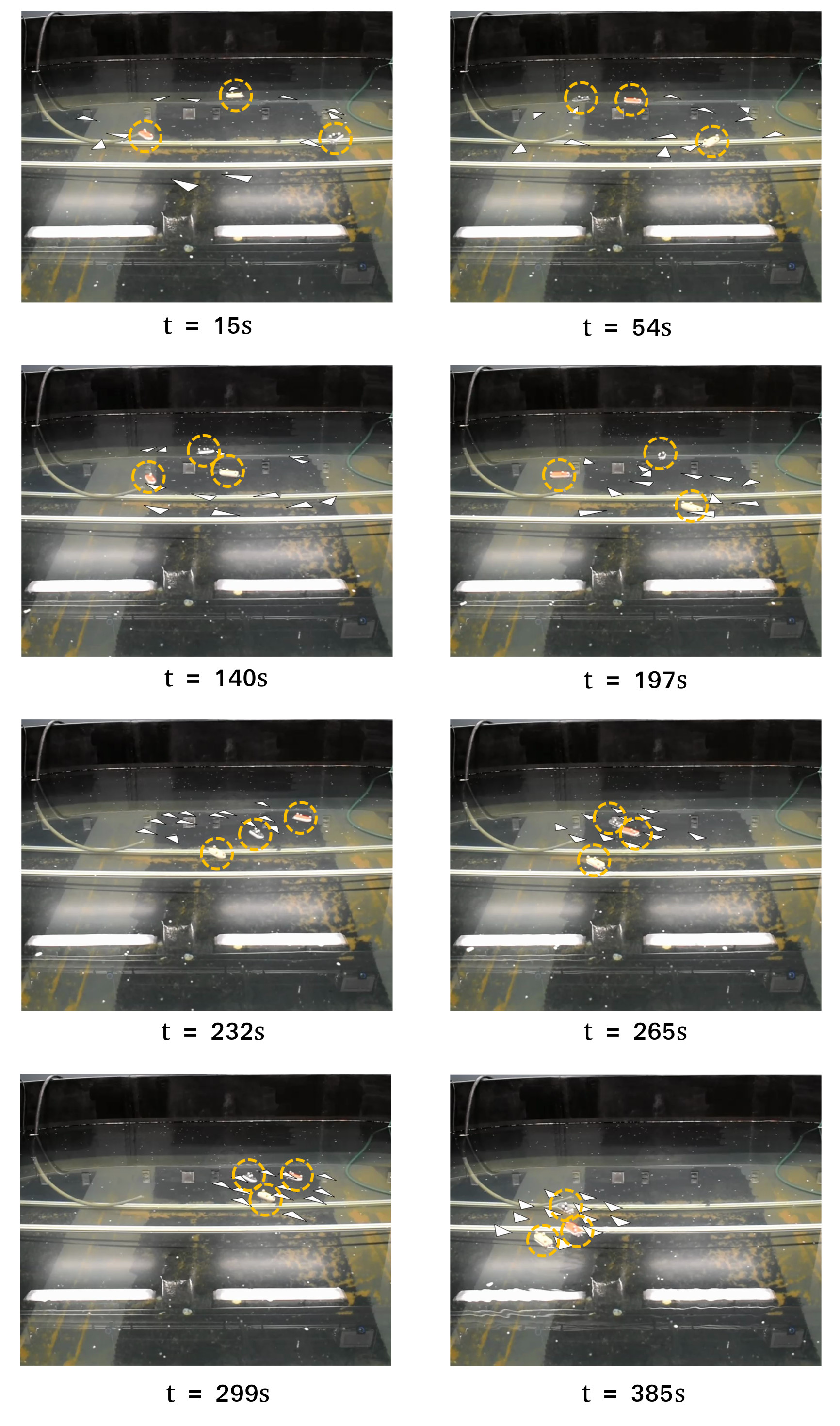}
\caption{Behavior of swarm with global sigmoidal repulsion during a ring to rotating transition. A delay of 40s was added at t = 60s.}
\label{fig:upenn_experiments_8_frames}
\end{figure}

The results from the experiments are presented in the following figures. 
Figure \ref{upenn_exp_type3rep_swarm_stats} depicts results from the translating to rotating swarm transition. 
The behavior of the swarm during the translating to ring swarm state transition
is observed in Figure \ref{upenn_exp_type2rep_swarm_stats}.
Finally, results for the transition between the ring to rotating swarm state 
are presented in Figure \ref{upenn_exp_type1rep_swarm_stats}, with Figure \ref{fig:upenn_experiments_8_frames} depicting the time series snapshots for $3$ ASV and 12 virtual agents transitioning from the ring to rotating swarm state. 

From the experimental results we observed and measured the different theoretically predicted swarm states. 
The ring swarm state can be identified by the low swarm polarization, center of mass acceleration, and center of mass speed indicative of an unaligned swarm that is stationary. 
The rotating swarm state can be identified by the high polarization, center of mass acceleration, and center of mass speed indicative of an aligned swarm that is constantly moving in a circle. 
All of these traits can be seen in Figure \ref{upenn_exp_type1rep_swarm_stats}.
The oscillations in the center of mass speed and acceleration, in Figure \ref{upenn_exp_type1rep_swarm_stats}, were the result of the swarm moving on an ellipsoidal trajectory, slowing down near the loci and speeding up near the semi-minor axis.
The dips in the polarization were the result of the existence of two groups of agents in the swarm that turn in different directions when the whole swarm reverses direction near the loci of the ellipse.  
The ellipsoidal motion gradually relaxed to a circle with all of the agents turning together in the same direction, with center of mass acceleration and velocity reaching steady state values and the polarization approaching a steady state value of 1.

Likewise, in Figure \ref{upenn_exp_type2rep_swarm_stats} there is a spike in polarization around the 50s mark.
This was due to the swarm reversing direction coherently before scattering into the ring swarm state.
In the translating and rotating swarm state, the ASVs adopted a hexagonal grid formation in the rough shape of a disk that remained rigid even during transitions between the rotating and translating state.

The experimental results successfully reproduced and extended the results obtained from the mixed reality UAV experiments described in section \ref{sec:uav_mr}. The ASV mixed reality experiments showed the persistence of the swarm states and the transition between the states even when collision avoidance routines were executed on all robots and virtual agents.
  These results are a step towards experimental validation because full scale robot experiments will require collision avoidance for safe robot-robot interactions.

\begin{figure}
  \centering
  \includegraphics[width=\linewidth]{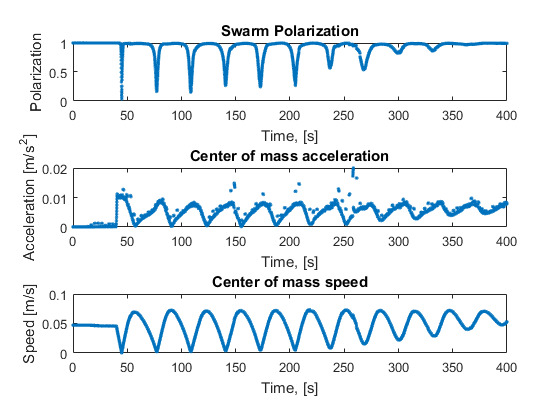}
  \caption{
    Swarm polarization, center of mass acceleration, and center of mass speed for ASV swarm using sensed sigmoidal repulsion\cite{repulsion_details} during a mixed reality experiments.
    A delay of 35s was introduced at t = 40s causing a translating to rotating transition.
    The dips in swarm polarization were caused by the agents disagreeing on which way to turn in the early stages of the transition, where the agents were in a degenerate rotating state (one characterized by the swarm moving back and forth along a line).
    As the eccentricity of the ellipse lessened, more agents agreed on which direction to turn at the vertices of the ellipse.}
  \label{upenn_exp_type3rep_swarm_stats}
\end{figure}

\section{Discussion}\label{sec:discussion}
Mixed reality provided a general framework to test the theoretical predicted behaviors with two different robotic platforms.
Our results cover a large swath of the parameter space for the theoretically predicted swarming patterns.
Our experimental results were obtained using vehicles with distinct dynamics, communication delays, and coupling strengths, nevertheless the experiments all exhibited the global patterns and pattern switches predicted by theory.

Different bifurcation diagrams were built for each combination of $\beta$ and $v_g$, which helped inform where the transition points between swarm states might occur in a physical experiment. 
The robotic platform differences contributed to a wide range of parameters that were necessary to achieve the different swarm states.
Additionally, different initial conditions for the swarms were outlined in the experimental section to compensate for platform differences.

In general, the UAV mixed reality experiments had faster dynamics and limited battery life, which restricted flight time.
Shorter UAV experiments tested individual behaviors at higher speeds, as well as select long experiments to observe transitions through all the swarm states.
Mixed reality allowed for the use of only one physical robot to capture complex interactions with the real world, while still maintaining a swarm of simulated agents \cite{new_results}.
Likewise, the 3 dimensional experiments with the dynamics of each robot constrained to the 2 dimensional slices demonstrate that the behaviors are possible for a UAV and a direction of future work is to consider what unconstrained 3 dimensional dynamics looks like with the addition of multiple real UAVs.
In the ASV mixed reality experiments, the dynamics were slower and the workspace constrained to 2D.
Nevertheless, these experiments captured the full effects of on-board collision avoidance routines on swarm pattern formation and pattern transitions and proved the persistence of the swarm states and the robustness of the communication delay induced switches in the global pattern formation.
These experiments are a first attempt at validating the theoretically derived results, Figure \ref{fig:dim_bifurcation}, in the presence of realistic real-world robot-robot and robot-environment interactions. 

During experimentation, there were problems that each robotic platform exhibited. 
In mixed reality experiments with the UAV for conditions testing the rotating swarm state, the UAV lagged behind the center of mass. 
This issue stems from internal safety features on the UAV to prevent high speeds. 
In the rotating swarm state the swarms acceleration is high, meaning that the simulated robots exceed the set point velocity $v_g$.
However, the real robot has internal mechanisms onboard which prevent rapid increases in speed.
We attempted to use different $\alpha$ values for the simulated robots and the real robot to try and pull the swarm together, but the real robot did not merge with the rest of the virtual cluster.
This behavior was not expressed in the ASV experiments.

\begin{figure}
  \centering
  \includegraphics[width=0.8\linewidth]{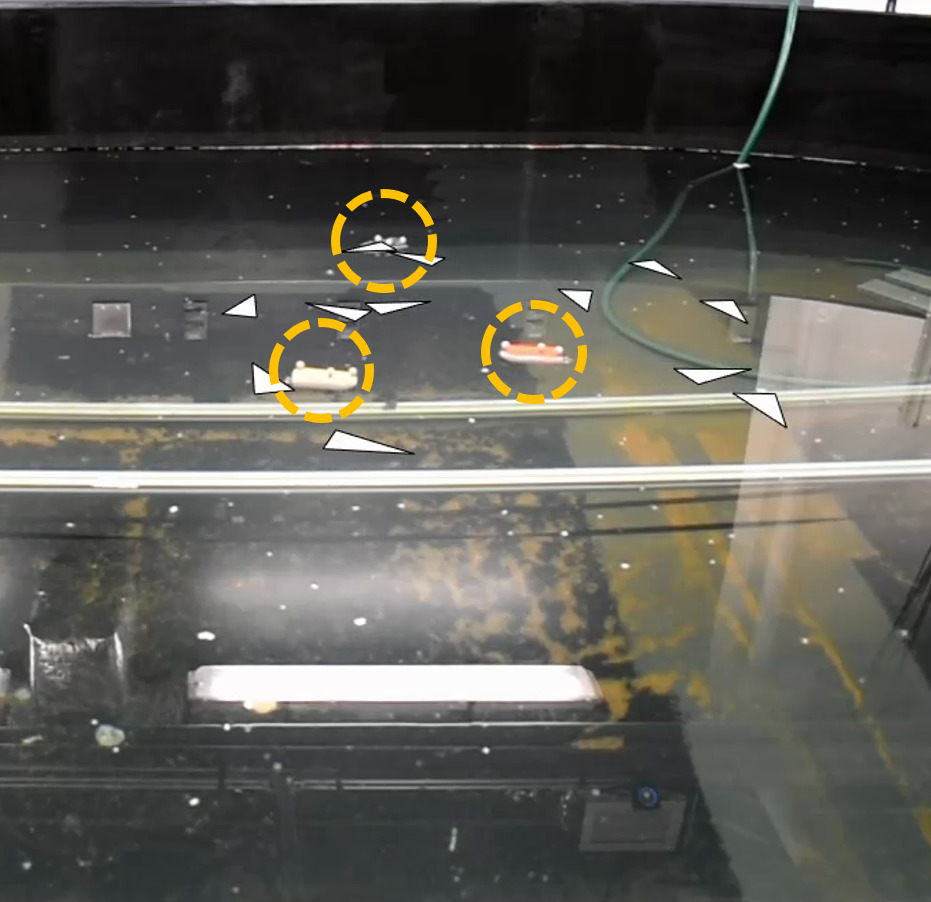}
  \caption{
  Example of ring state with occupied middle region due to the slow acceleration of the ASVs. The inability of the ASVs to accelerate back up to full speed after a run-in with another agent resulted in them traveling on a highly eccentric path that frequently passed through the center.}
  \label{upenn_exp_loose_ring}
\end{figure}

The ASV mixed reality experiments introduced collision avoidance to the swarm dynamics.
The addition of collision avoidance risks the complete destruction of the theoretically predicted swarming patterns. 
Since collision avoidance presents a hard constraint, pairs of vehicles that are in immediate danger of collision would end up being forcefully pushed apart by their collision avoidance routines. 
The result is either rings of larger thickness or agents that swing towards the center of the ring, Figure \ref{upenn_exp_loose_ring}.

\begin{figure}
    \subfigure[ $\tau_0 = 0.01$\,s, $\Delta \tau = 0.3$, update $\tau$ every $10$\,s ]{
  \includegraphics[width=\linewidth]{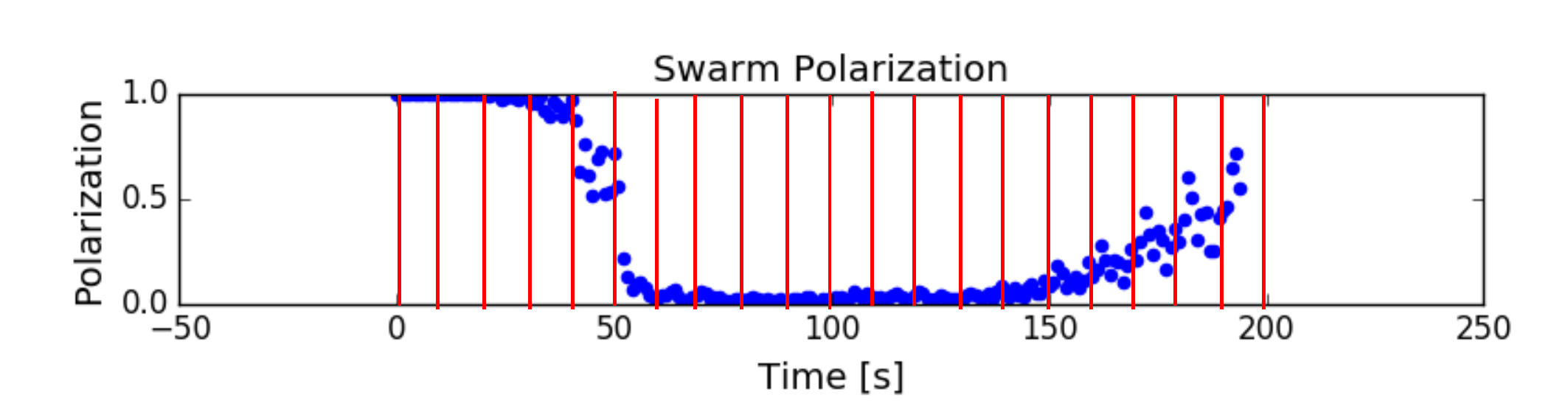}
    \label{fig:mult_stable_fast}
  }
  \subfigure[ $\tau_0 = 3.0$\,s and $\Delta \tau = 0.5$, update $\tau$ every $50$\,s ]{
  \includegraphics[width=\linewidth]{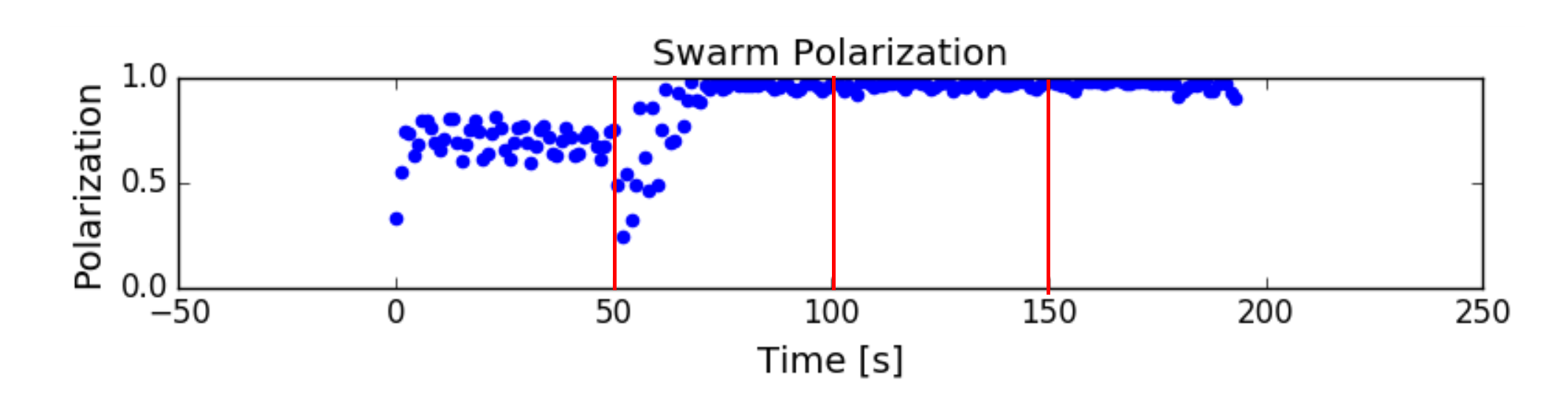}
    \label{fig:mult_stable_medium}
  }
  \subfigure[ $\tau_0 = 2.5$\,s and $\Delta \tau = 1.0$, update $\tau$ every $75$\,s ]{
  \includegraphics[width=\linewidth]{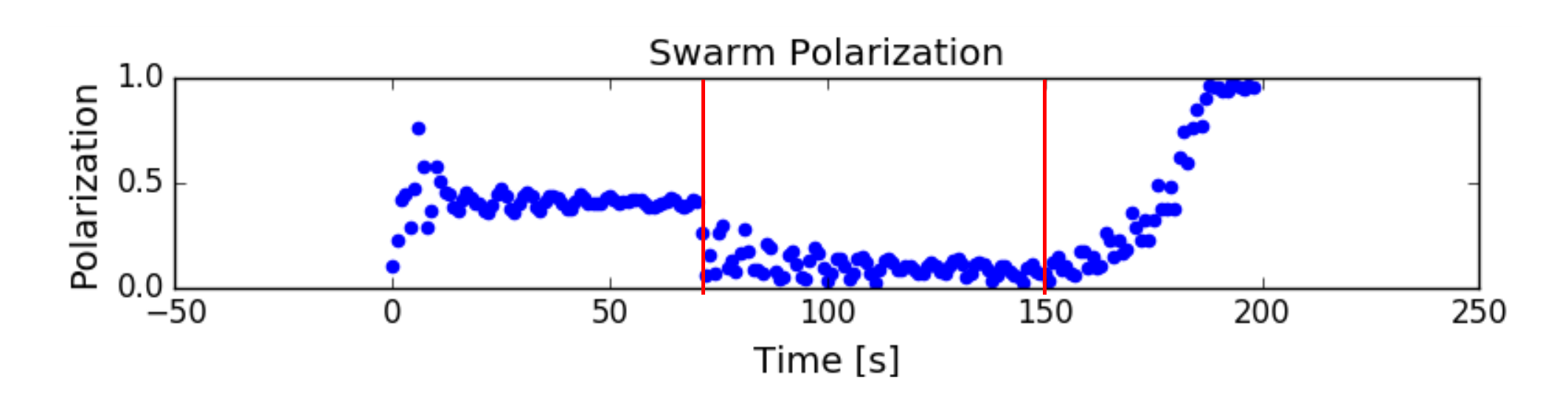}
    \label{fig:mult_stable_slow}
  }
  \caption{
    All graphs represent a 200\,s experiment, with 1 UAV, and 49 simulated robots.
    The parameters used in both experiments were: $\alpha = 1.0$\, 1/s$^2$, $\beta = 20.0$\,m/s$^2$, and $v_g = 0.2$\,m/s.
    In these experiments $\tau$ started at different initial points and was updated at different times throughout the experiment.
    Figure \ref{fig:mult_stable_fast} is an experiment that started in a translating state, transitioned to the ring state, and around 150\,s, or $\tau = 4.81$ began to transition to the rotating state. 
    Figure \ref{fig:mult_stable_medium} is an experiment that started in the ring state and at around 50\,s, or $\tau = 3.5$\,s transitioned to the rotating state. 
    Figure \ref{fig:mult_stable_slow} is an experiment that started in the ring state and at around 175\,s, or $\tau = 4.0$\,s began to transitioned to the rotating state. 
    These three plots highlight that a transition from the ring to rotating state can occur with different initial conditions and different amounts of time with a certain amount of delay in communication.
    This is due to the multi-stability of the patterns along the bifurcation curve. 
  }
  \label{multi_stable_figure}
\end{figure}

It is hard to analytically describe the multi-stability present in Equation \ref{eq:dimensional}.
In addition to multi-stability observed in Figure \ref{bifurcation_data_figure}, multi-stability was also observed in Figure \ref{multi_stable_figure}, where given three similar initial conditions the amount of time the system used a specific amount of delay impacted the transition between the ring and rotating swarm state.
For example, in Figure \ref{fig:mult_stable_fast} the transition to the rotating swarm state happened for $\tau = 4.81$\,s, in Figure \ref{fig:mult_stable_medium} the transition happened for $\tau = 3.5$\,s and in Figure \ref{fig:mult_stable_slow} the transition happened at $\tau = 4.0$\,s.
These variances in transition points come from different initial configurations of the swarm, and the evolution of the system.


\section{Conclusion and Future Work}
\label{sec:conclusion}
Through manipulating communication delay and coupling strength we took the first significant steps to test different emergent swarm states using mixed reality.  
The mixed reality framework allowed for the study of emergent swarm behavior through the use of simulation to increase the number of agents while maintaining critical real world interactions, which are hard to model, with physical platforms. 
Because mixed reality can handle many different levels of abstraction we were able to use two distinct robotic platforms to validate the swarm behavior.
We selected the level of abstraction necessary for the behavior to exhibit, and in our case it was focused on agents that exhibited simple dynamics and used delayed information as would be done in the real world. 
Both the ASVs and the UAV tested all three of the theoretically predicted behavior along with transitions between the predicted swarm states. 
This emphasizes that the proposed swarm model has the potential to be applicable across platforms, and highlights the impacts of communication delay on systems behavior.

There is a range of theory that supports the proposed model, and the presented results are significant steps toward showing theory is valid, as well as also demonstrating that there are areas of interest that the theory is not capturing. 
Understanding the multi-stability in this system is difficult analytically, but with the use of simulation and mixed reality we were able to observe multi-stability and the impacts it has on the proposed theoretical model when paired with real vehicles.

The next steps for this work include investigating how the addition of more real world assumptions change the predicted emergent patterns.
For example, our communication model of global coupling is not practical with
all real robots due to network limitations. Next steps, are to study the impacts of range based communication.
While often swarms are studied for homogeneous agents it is also possible to consider the impacts of heterogeneity as was done in Szwaykowska et al. \cite{Szwaykowska2016}, this may require different types of collision avoidance to be used to account for different hardware limitations.
Finally, there exist different dynamic models, where delay can be added.
This presents new potential patterns to be studied using the described forms of analysis. 
These changes will continue to add to the understanding of our current models and the use of this type of emergent behavior in the physical world.

\section{Acknowledgment}
Edwards, Triandaf, Hindes, and Schwartz gratefully acknowledge the Office of Naval Research (ONR) for their support under N0001412WX20083, support of the NRL Base Research Program N0001412WX30002, and the NRL Karles Fellowship Program.
DeZonia and Hsieh gratefully acknowledge the support of ONR Award No. N00014-18-1-2580 and ARL DCIST CRA W911NF-17-2-0181.

\section{Appendix}
\subsection{Mean Field Analysis}
We will consider the mean field analysis when the collision avoidance term is anti-symmetric to the robot's neighbors, resulting in negligible impact to the remaining analysis. 
We will consider the mean of the swarm as $R(t) = \frac{1}{N} \sum_{i=0}^N r_i(t)$, which is the average center of mass, for all agents in the swarm. 
$R(t)$ has units meters. Likewise consider that $\delta r_i(t) = r_i(t) - R(t)$ is the distance of each particle $r_i$ in meters from the center of mass in meters, thus $\delta r_i(t)$ has units meters.

We know that the $\sum_{i=0}^N \delta r_i = \sum_{i=0}^N \delta \dot{r_i} = \sum_{i=0}^N \delta \ddot{r_i} = 0$. 
We will now consider the substitution into Equation \ref{eq:dimensional} using the above relationships:

\begin{align}
\ddot{R} + \ddot{\delta r_i} = \beta(v_g^2 - || \dot{R} + \delta \dot{r_i}||^2) (\dot{R} + \delta \dot{r_i}) \nonumber \\
- \frac{\alpha}{N} \sum_{j=1}^N (R + \delta r_i - R^{\tau} - \delta r_j^{\tau}), 
\end{align}

where $R^{\tau} = R(t - \tau)R$ and the units of $\ddot{R}$ is m/sec$^2$.
Multiplying out all the relationships and condensing like terms we get the following:

\begin{align}
  \ddot{R} = -\ddot{\delta r_i} + \beta \dot{R} (v_g^2 - ||\dot{R}||^2)  + \beta  (v_g^2 - || \dot{R}||^2) \delta \dot{r_i}
  \nonumber \\
  - \beta ( || \delta \dot{r_i}||^2 + 2<\dot{R}, \delta \dot{r_i}>)(\dot{R} + \delta \dot{r_i})  
  \nonumber \\
  - \frac{\alpha}{N} \sum_{j=1}^N (R - \delta r_i - R^{\tau} - \delta r_j^{\tau}),
\end{align}

where $<.,.>$ is the dot product.
Taking the sum over all $r_i$ in the swarm and dividing by $N$ you get the following equation:

\begin{align}
  \ddot{R} = -\frac{1}{N}\sum_{i=0}^N \delta \ddot{r_i}   + \beta \dot{R} (v_g^2 - ||\dot{R}||^2) 
  + \frac{\beta}{N} (v_g^2 - || \dot{R}||^2)\sum_{i=0}^N\delta \dot{r_i} 
  \nonumber \\
  - \frac{\beta}{N} (|| \delta \dot{r_i}||^2 + 2 < \dot{R}, \delta \dot{r_i}>)(\dot{R} + \delta \dot{r_i})
  \nonumber \\
  - \frac{\alpha}{N}(R - R^{\tau}) - \sum_{i=0}^N\sum_{j=0, i \neq j}^N (\delta r_i - \delta r_j^{\tau}).
\end{align}

The observations were made in previous work that the tendency is for particles to stay close to the center of mass, which means the impacts of deviations from the center of mass play less and less importance as N increases \cite{Szwaykowska2016}. If we consider as $N \rightarrow \infty$ then we can ignore the individual impacts of $\delta r_i$. We know that $\sum_{i=0}^N \delta \ddot{r_i} = 0$, eliminating the first term in the equation.

Thus giving us the following equation in m/s$^2$:

\begin{align}
\ddot{R} = \beta(v_g^2 - ||\dot{R}||^2) \dot{R} - \frac{\alpha}{N}(R - R^{\tau}).
\end{align}
    
\subsection{Evaluation Metrics}
The mathematics behind the prediction of the radius of the ring and rotating states along with the derivation of the angular velocity, was originally presented in Szwaykowska et al. \cite{Szwaykowska2016}.
$\rho_{ring}$ is the radius of the ring and $\omega_{ring}$ is the angular velocity of the swarm. The equations are:

\begin{align}
  \rho_{ring} = \frac{\sqrt{ 1 / a }}{v_g \beta},
  \\
  \omega_{ring} = \frac{\sqrt{a}}{v_g^2 \beta}. 
\end{align}

Likewise, the radius of rotation and angular velocity of the swarm in the rotating state are represented by $\rho_{rot}$ and $\omega_{rot}$.
The equations are:
\begin{align}
  \rho_{rot} = \frac{1}{|\omega_{rot}|v_g \beta}\sqrt{1 - \frac{a\sin(\omega_{rot} \tau)}{\omega_{rot}} },
  \\
  \omega_{rot}^2 = \frac{a}{v_g^2 \beta} [ 1 - \cos(\omega_{rot} \tau)]. 
\end{align}

Note that the $\alpha$ used in the experiment needs to be converted to $a$ using the following conversion: $ a = \frac{\alpha}{\beta^2 v_g^4}$.
Due to noise related to measuring $\omega_{ring}$ and $\omega_{rot}$ values from experimental data, we instead compute the period of the swarm in the ring and rotating swarm state through the following conversion $T_{ring} = \frac{2\pi }{\omega_{ring}}$ and $T_{rot} = \frac{2\pi}{\omega_{rot}}$.

\begin{thebibliography}{58}
\bibitem [{rep()}]{repulsion_details}%
  \BibitemOpen
  \href@noop {} {}\bibinfo {howpublished} {Three different variants of sigmoid
  repulsion were tested, where current robot position data was used in each
  case. The first type was global sigmoid repulsion, where all agents had
  delayed attraction and an infinite range of repulsion. The next type was
  local sigmoid repulsion, where all agents had delayed attraction and a finite
  range of repulsion. Finally there was sensed sigmoid repulsion, where all
  agents had delayed attraction outside a simulated local sensing radius and
  up-to-date information within that radius with finite range repulsion. Sensed
  sigmoid repulsion attempts to emulate a realistic interaction model where
  agents make full use of their sensors by fusing communicated and sensed
  position data about other agents. Four experiments were run using each type
  of repulsion: global, local, and sensed sigmoid repulsion. The collision
  avoidance parameters used were $c_{r} = 0.0067$\,m/s$^2$, $R_{rep} = 0.1$\,m,
  and $k = 40$.}\BibitemShut {Stop}%
\bibitem [{new()}]{new_results}%
  \BibitemOpen
  \href@noop {} {}\bibinfo {howpublished} {The size of the AscTec Pelican UAV
  limited the number of robots we could deploy safely, this lead to using the
  Crazyflie Micro-quadrotor. These robots are 27g and have 7 minutes of flight
  time. We have seen successful ring swarm state experiments with 8 real robots
  and 16 simulated robots. Further testing that we are using mixed reality as
  steps toward validating theoretical models.}\BibitemShut {Stop}%
\end{thebibliography}
%

\end{document}